\documentclass[twocolumn,showpacs,preprintnumbers,superscriptaddress]{revtex4}

\usepackage{amsfonts,amssymb,amsmath,amsthm}

\advance\topmargin 1.0cm

\usepackage{dcolumn}
\usepackage{bm}

\usepackage{color}
\usepackage{epsfig}

\newcommand{\ncd}{\newcommand}

\ncd{\QC}{$\mbox{QC}_{\cal{C}}\;$}
\ncd{\QCpr}{${\mbox{QC}_{\cal{C}}}^\prime\;$}
\ncd{\QCns}{$\mbox{QC}_{\cal{C}}$}
\ncd{\QCprns}{${\mbox{QC}_{\cal{C}}}^\prime$}
\ncd{\ds}{\displaystyle}
\ncd{\ovl}{\overline}
\ncd{\iden}{1 \hspace{-1.0mm}  {\bf l}}
\ncd{\im}{\text{i}}

\begin{document}

\title{A quantum cellular automaton for universal quantum computation}

\author{Robert Raussendorf}
\affiliation{California Institute of Technology,\\
Institute for Quantum Information, Pasadena, CA 91125, USA}

\date{\today}

\begin{abstract}
  I describe a quantum cellular automaton capable of performing universal 
  quantum computation. The automaton has an elementary transition 
  function that acts on Margolus cells of $2\times 2$ qubits, and both the 
  ``quantum input'' and the program are encoded in the initial state of the 
  system.
\end{abstract}

\pacs{3.67.Lx, 3.67.-a}

\maketitle


\section{Introduction}
\label{intro}

A number of physical systems that are considered for the realization of a universal quantum computer, such as optical lattices \cite{BA} or arrays of micro-lenses \cite{BE}, possess a translation symmetry in the arrangement of qubits and their mutual interaction. Quantum cellular automata (QCA) represent a suitable framework to explore the computational power of such physical systems, because they respect this symmetry. A priori, translation invariance may be regarded as a severe limitation since it constrains the degree of control that can be exerted to the quantum system. However, it has been demonstrated that one-dimensional QCA can efficiently simulate any quantum Turing machine \cite{WTR}. 

Further it has been shown that there exists a universal QCA which can simulate any other automaton with linear slowdown \cite{WvD}, and that every reversible QCA can be represented in a generalized Margolus partitioning scheme \cite{SW}. Proposals with an emphasis on experimental viability have outlined how generic physical systems can be used as quantum computation devices if equipped with a minimal amount of external control. Among the described mechanisms are global control via sequences of resonant light pulses \cite{SL} or modulation of a coupling constant \cite{SB1,Lev}, and individual control over one of the elementary cells \cite{SB2}.

At this point, one may abandon all algorithm-specific control during the process of computation and ask ``How intricate do quantum cellular automata have to be such that they can perform useful tasks in quantum information processing?''. A quick answer may be ''Simple, by definition.'' However, when QCAs are tuned for algorithmic application, it may occur that---while the simple composition is retained---the elementary cells and neighborhood schemes become complicated. An interesting facet of the answer to the above question has been provided in \cite{Brenn}, where a very simple QCA for quantum data transmission has been devised (also see \cite{SW}). Motivated by a recent result \cite{Wotz}, where universal computation via autonomous evolution of a 10-local Hamiltonian is described, here I consider quantum computation in the cellular automaton scenario. I explicitly construct a computationally universal two-dimensional QCA whose transition rule is based on a four-qubit unitary. 

\section{Construction of the automaton} 

Consider a two dimensional lattice of size $2s \times 2r$ with periodic 
boundary conditions, i.e. a torus. Each lattice site carries a qubit. 
The transition rule for the QCA is described in terms of a Margolus 
partitioning \cite{Marg}. The lattice is partitioned into cells of size 
$2\times 2$, and there is a separate partitioning for the time $t$ being 
even or odd, respectively. One may choose a coordinate system on the torus 
with 
axes parallel to what were the boundaries before identification. For $t$ even, the qubits in the upper left corner of each cell have both coordinates even, and for $t$ odd they have both coordinates odd. Thus, a cell in step $t$ overlaps  with four cells of step $t-1$. 

The transition of the QCA from time $t$ to $t+1$ proceeds by simultaneously 
applying a unitary transition function $\tau$ to each cell. 
For a suitable choice of the 4-qubit unitary $\tau$ one can 
perform universal quantum computation with the described QCA. Specifically, a quantum logic network of local and next-neighbor unitary gates with width $2s$ and depth $r$ can be simulated. 

\begin{figure}[ht]
  \begin{center}
    \epsfig{file=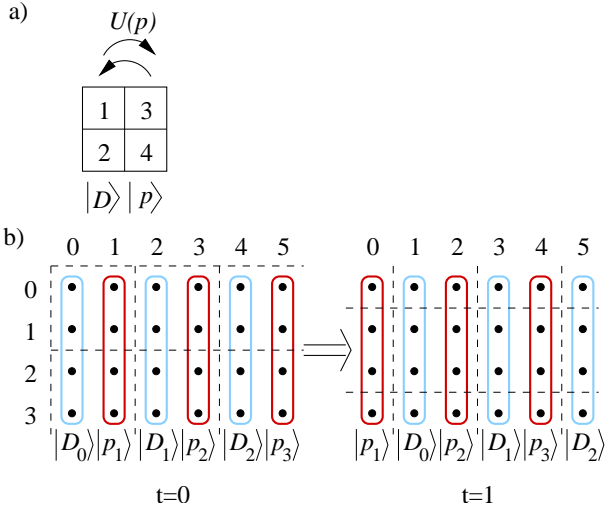, width=8cm}
    \caption{\label{cells}Transition of the QCA from $t$ to $t+1$. a: Margolus 
      cell of $2\times2$ qubits, with the part $|D\rangle_{12}$ of a data- and 
      $|p\rangle_{34}$ of the program column.  
      b: Before and after the first transition. The program columns move left 
      and the data columns move right. The dashed lines indicate the 
      partitioning into Margolus cells.}
  \end{center}
\end{figure}

With a labeling of particles as illustrated in Fig.~\ref{cells}a, the following elementary transition function is chosen:
\begin{equation}
  \label{elem}
  \begin{array}{rcl}
    \tau &=& \displaystyle{\text{S}(1,3)\,\text{S}(2,4)\, H_1 
      \exp\left( -\im \frac{\pi}{8} \frac{1-Z_3}{2}Z_1\right)}\\
    && \displaystyle{\exp\left(\im \pi \frac{1-Z_4}{2} \frac{1-Z_1}{2} 
      \frac{1-Z_2}{2} \right).}
  \end{array}
\end{equation}
Therein, $\text{S}(a,b)$ denotes a SWAP-gate between qubit $a$ and $b$, $H_1$ is a Hadamard transformation on qubit $1$, and $Z_c$ denotes a Pauli phase flip operator applied to qubit $c$. 

If $|p\rangle_{34}$ is a state in the computational basis, it effectively stores two classical bits, $p^{(3)}$ and $p^{(4)}$. Then, the transition function $\tau$ amounts to a classically controlled unitary operation $U(p)$ applied to $|D\rangle_{12}$. The bit $p^{(4)}$ triggers a $\Lambda(Z)$ gate applied to $|D\rangle_{12}$, and $p^{(3)}$ a $\pi/4$-phase gate $\exp(-\im \pi/8\, Z_1)$ applied to qubit 1 of $|D\rangle_{12}$. In this way, $|p\rangle$ encodes an elementary step of a program, carried out on the two-qubit ``data'' $|D\rangle_{12}$. The SWAP-gates allow the quantum data and the program to pass by another, such that $|D\rangle_{12}$ can interact with subsequent program registers.  

\begin{figure}
  \begin{center}
    \epsfig{file=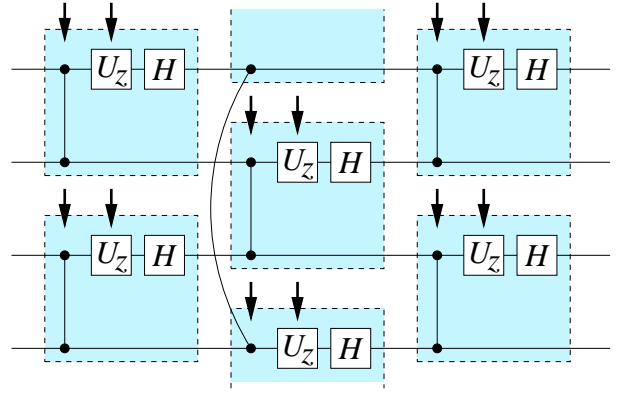, width=8cm}
    \caption{\label{qln}The quantum logic network simulated by the QCA. 
      The gates in each shaded box result from one application of the 
      elementary transition function. The arrows denote classically 
      controlled gates which are triggered by the program registers.}
  \end{center}
\end{figure}

The sets of qubits on the torus with the same first coordinate $x$, $0\leq x \leq 2r-1$, are called columns. At time $t=0$ all even columns contain data registers $|D_i\rangle$, and all odd columns contain program registers $|p_j\rangle$. The initial state of the automaton is
\begin{equation}
  \label{init}
  |\psi(0)\rangle = \bigotimes_{i=0}^{r-1}|D_i(0)\rangle_{2i} |p_{i+1}
  \rangle_{2i+1}.
\end{equation}
Here and in the following the labels inside the kets specify the state and the ones outside specify the location of the support within the lattice i.e., the column. For example, $|D_1(0)\rangle_2$ is data register No. 1 at time $t=0$, located on column 2. Of all data registers only $|D_0\rangle$ is used, the others are auxiliary. When the QCA starts to run, the data registers move right (counter-clockwise, as seen from top) and the program registers move left (clockwise), by one column in each time step. When passing the data registers, the program registers $|p\rangle$ control unitary transformations $U(p)$ acting upon the data registers $|D\rangle$. In this way, a program specified by the data $p_1,...,p_r$, with $p_1$ encoding the first and $p_r$ the last step, is carried out on the quantum data register $|D_0\rangle$. The program that is carried out corresponds to a quantum logic network of local and next-neighbor gates in a particular arrangement; see Fig.~\ref{qln}. Such networks are sufficient for universal quantum computation, as is discussed in detail further below. The same program steps that are applied to the register $|D_0\rangle$ are also carried out on the auxiliary data registers $|D_i\rangle$, $1 \leq i \leq r-1$, but in scrambled order. Therefore, these registers are not used. 

In course of computation, both data and program travel across half the torus. 
When the automaton has run for $r$ time steps, the computation is finished and the register $|D_0(r)\rangle_r$ can be read out from column $r$, via local measurements.\medskip

Let $\tau_{i,j}$ denote the elementary transition function (\ref{elem}) applied to the cell $(i,j)$. Therein, $i$ is the column coordinate of the upper left qubit in the cell, and $j$ the respective coordinate within the column. Then, the unitary transition function $T_i$, acting upon two consecutive columns $i$, $i+1$, is 
\begin{equation}
  T_i=\bigotimes_{j=0}^{s-1} \tau_{i,[2j+i]_{2s}}.
\end{equation}
Therein, $[2j+i]_{2s}$ is a shorthand for $2j+i\; \text{mod} \; 2s$ that will be used throughout the remainder of the paper. 

If $|p\rangle$ is a state in the computational basis, then
\begin{equation}
  \label{ProgStep}
  T_i\, \left(|D\rangle_i \otimes |p\rangle_{i+1}\right) = |p\rangle_i \otimes 
  \left( U(p)\, |D\rangle \right)_{i+1}.
\end{equation}
Therein, $U(p)$ is a unitary transformation chosen by $p$ containing a Hadamard- and possibly a $\Lambda(Z)$- and a $\pi/4$-phase gate, in accordance with (\ref{elem}).

The global transition function $T: |\psi(t)\rangle \longrightarrow |\psi(t+1)\rangle$ is, for even $t$ given by $T_\text{e}= \bigotimes_{i=0}^{r-1} T_{2i}$, and for odd $t$ by $T_\text{o}=\bigotimes_{i=0}^{r-1} T_{2i+1}$. In both cases it can be written in the form
\begin{equation}
  \label{global}
  T = \bigotimes_{i=0}^{r-1} T_{[2i+t]_{2r}}.
\end{equation}
Now, the state $|\psi(t)\rangle$ of the QCA at time $t$ is
\begin{equation}
  \label{evol}
  |\psi(t)\rangle = \bigotimes_{i=0}^{r-1} |D_i(t)\rangle_{[2i+t]_{2r}} 
  \otimes |p_{[i+t+1]_r}\rangle_{[2i+t+1]_{2r}},
\end{equation}
with the data register $i$ at time $t$ given by
\begin{equation}
  \label{Dit}
  |D_i(t)\rangle = \left[\prod_{k=1}^t U(p_{[i+k]_r}) \right] |D_i(0)\rangle.
\end{equation}
The unitaries $U(p_{[i+k]_r})$ are ordered in ascending order with $k$, i.e. $U(p_{[i+1]_r})$ acts first. For $i=0$ and $t=r$, in particular, one finds that 
\begin{equation}
  |D_0(r)\rangle = \left[\prod_{k=1}^r U(p_k) \right] |D_0(0)\rangle
\end{equation}
is the output quantum register, with the unitaries $U(p_1) ... U(p_r)$ 
applied in the correct order to the quantum register in its input state, 
$|D_0(0)\rangle$.

Before proving (\ref{evol}), let us recover therein some features of the QCA that were stated before. It is easy to see that---apart from being moved---the program registers remain unchanged throughout the evolution, and that there is no entanglement across columns. The data registers indeed move right, and the program registers  left (with $i+t+1=i^\prime$, $|p_{[i+t+1]_r}\rangle_{[2i+t+1]_{2r}} = |p_{[i^\prime]_r}\rangle_{[2i^\prime-t-1]_r}$). After $r$ time steps, the output quantum register $|D_0(r)\rangle$ can be read out from column $r$.

Eq. (\ref{evol}) is proved by induction. First note that for $t=0$ it reduces to (\ref{init}). Further,
\begin{eqnarray}
  T|\psi(t)\rangle &=& \bigotimes_{i=0}^{r-1} T_{[2i+t]_{2r}} 
  |D_i(t)\rangle_{[2i+t]_{2r}} |p_{[i+t+1]_r}\rangle_{[2i+t+1]_{2r}} \nonumber 
  \\
  &=& \bigotimes_{i=0}^{r-1} \left( U(p_{[i+t+1]_r}) 
  |D_i(t)\rangle\right)_{[2i+t+1]_{2r}} \otimes \nonumber \\
  && \;\;\;\;\;\;\;  |p_{[i+t+1]_r}\rangle_{[2i+t]_{2r}} \nonumber \\
  &=& \bigotimes_{i=0}^{r-1} {|D_i(t+1)\rangle}_{[2i+(t+1)]_{2r}} \otimes 
  \nonumber \\
  &&  \;\;\;\;\;\;\; |p_{[i+(t+1)+1]_r}\rangle_{[2i+(t+1)+1]_{2r}} \nonumber \\
  &=& \displaystyle{|\psi(t+1)\rangle.}
\end{eqnarray}
Here, the first line follows by the definitions of $T$ and $|\psi(t)\rangle$, (\ref{global}) and (\ref{evol}), and the second follows by (\ref{ProgStep}). 
The 
third line follows by (\ref{Dit}) and, for the $|p\rangle$-part, the substitution $i \rightarrow i+1$ under which the product is invariant. 
$\Box$ \medskip

\begin{figure}
  \begin{center}
    \epsfig{file=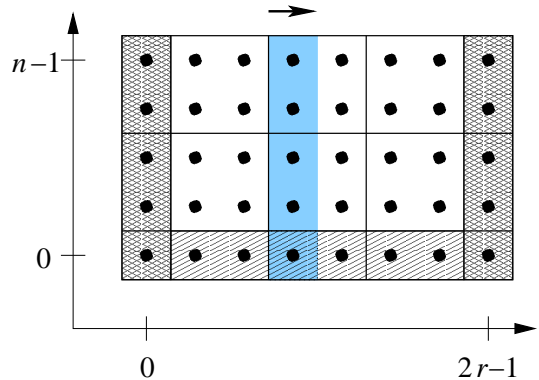, width=7cm}
    \caption{\label{boundary}Boundary specifications for the described QCA on 
      a planar sheet. The $2\times 2$-cells displayed in white are acted upon 
      by $\tau$, as usual. To the hatched $1\times 2$-cells a SWAP-gate is 
      applied, and to the cross-hatched cells the identity operation. The 
      column in gray underlay represents a data register moving right.}
  \end{center}
\end{figure}

Finally, it is shown that the quantum logic network simulated by the described 
QCA is indeed universal, as stated. The CNOT-, the Hadamard- and the $\pi/4$-phase gate $\exp(-\im \pi/8 \,Z)$ form a universal gate set \cite{Boy}. For the described QCA, one can independently apply the Hadamard-, the $\pi/4$-phase gate and the identity to the simulated logical qubits, by choosing the following sequences of $p^{(3)}$-bits
\begin{equation}
  {\bf{p}}^{(3)} = \left\{
  \begin{array}{rl}  
    0\, 000\, 000\, 000, & \mbox{for }U=I,\\
    0\,101\,101\,101, &  \mbox{for }U=H,\\
    1\, 000\, 000\, 000, & \mbox{for }U=U_z[\pi/4].
  \end{array} 
  \right.
\end{equation}
with all $p^{(4)}$-bits zero. In the way they are constructed here (no claim that this is any close to optimal), the one-qubit operations from the universal set require 10 successive applications of $p^{(3)}$-controlled gates and thus 20 time steps of the QCA. A next-neighbor CNOT-gate that acts within this cycle can be constructed from a $\Lambda(Z)$- and two Hadamard-gates. The long-distance CNOT-gates may then be constructed with the help of next-neighbor SWAP-gates, which themselves consist of three next-neighbor CNOT-gates. This completes our construction of a QCA capable of performing universal quantum computation.\medskip

\noindent
Two remarks: 1) The described QCA may, with some right, be called a deterministic programmable quantum gate array, but this notion is already in use for a construction that has been proven not to exist \cite{Nie}. Our QCA is consistent with this result. The program information is classical and all program states $\bigotimes_{i=1}^r |p_i\rangle_{2i-1}$ are orthogonal, as required in \cite{Nie}. Further, from the viewpoint of temporal complexity, the described QCA is---within a constant---as efficient as a quantum logic network with local and next-neighbor gates.

2) That the described QCA lives on a torus simplifies the discussion, but is not essential. A planar sheet of size $n\times 2r$ is sufficient for simulation of the discussed networks of $r$ time steps and $n$ qubits. Fig.~\ref{boundary} specifies the operations on the boundary that differ from $\tau$. When this modified QCA is run, in the bulk the data registers still move right and the program registers left. On the left and right boundary, however, the registers are reflected. As a consequence, on the left side of the lattice, reflected program registers are acted upon by left-moving program registers as if they were data. More severely, on the right side of the cluster, reflected data registers act upon right-moving data registers as program. Therefore, the state of the QCA is no longer a tensor product of the column states, but instead an entangled state supported by many columns grows from the right, by one column in each time step. However, none of that has an impact on the data register $|D_0\rangle$, which---as before---can be read out from column $r$ after $r$ transitions of the automaton.

\section{Conclusion}

I have described a quantum cellular automaton for universal quantum computation. The  transition function from one time-step to the next is generated by a four-qubit unitary transformation acting on Margolus cells of size $2\times 2$.  The program is encoded in the initial state of the system, and the automaton is left to its autonomous evolution from initialization to readout.

Coming back to our initially posed question, it is found that QCAs performing complex tasks in quantum information processing can indeed be constructed for compact cells and neighborhood schemes.

\begin{acknowledgments}

I would like to thank Pawel Wocjan and Sergey Bravyi for discussions.
This work was supported by the National Science Foundation
under grant number EIA-0086038.
\end{acknowledgments}


\end{document}